\documentclass[aps,prr,twocolumn,showpacs,superscriptaddress,amsmath,amssymb,floatfix,10pt,reprint]{revtex4-1}  
\usepackage{graphicx}  
\usepackage{dcolumn}   
\usepackage{bm}        
\usepackage{amssymb}   
\usepackage{framed}
\usepackage{amsmath}
\usepackage{hhline}
\usepackage{subfigure,amsmath,verbatim,moreverb}
\usepackage{tabularx}
\usepackage{lipsum}
\usepackage{longtable}
\usepackage{booktabs}
\usepackage{threeparttable}
\usepackage{booktabs,amssymb,siunitx}
\usepackage{latexsym}
\usepackage{dcolumn}
\usepackage{amsmath}
\usepackage{epsf}
\usepackage{float}
\usepackage{hyperref}
\usepackage{xcolor}
\usepackage{etoolbox}
\AtBeginEnvironment{align}{\setcounter{subeqn}{0}}
\newcounter{subeqn} %

\begin{document}

\title{A way of resolving the order-of-limit problem of Tao-Mo semilocal functional }
\author{Abhilash Patra}
\email{abhilashpatra@niser.ac.in}
\affiliation{School of Physical Sciences, National Institute of Science Education and Research, HBNI, 
Bhubaneswar 752050, India}
 \author{Subrata Jana}
 \email{subrata.jana@niser.ac.in, subrata.niser@gmail.com }
 \affiliation{School of Physical Sciences, National Institute of Science Education and Research, HBNI, 
 Bhubaneswar 752050, India}
\author{Prasanjit Samal}
\email{psamal@niser.ac.in}
\affiliation{School of Physical Sciences, National Institute of Science Education and Research, HBNI, 
Bhubaneswar 752050, India}

\date{\today}

\begin{abstract}

It is highlighted recently that the Tao-Mo (TM) [Phys. Rev. Lett. {\bf117}, 073001 (2016)] semilocal exchange-correlation
energy functional suffers from the order-of-limit problem, which affects the functional
performance for phase transition pressures [J. Chem. Phys. 152, 244112 (2020)]. The root of the
order-of-limit problem of the TM functional inherent within the interpolation function, which switches between the compact density and the
slowly varying density. In this paper, we propose a different switch function that avoids the order-of-limit problem and
interpolates correctly between the compact density and the slowly varying fourth-order density correction. 
By circumventing the order-of-limit problem, the proposed form enhances the applicability of the original TM functional on the diverse
nature of the solid-state properties. Our conclusion is ensured by examining the functional in predicting properties related to the 
general-purpose solids, quantum chemistry, and phase transition pressure. Besides, we reasonably discuss the 
connection between the order-of-limit problem, phase transition pressure, and band gap of solids.

\end{abstract}

\maketitle

\section{Introduction}

The Kohn-Sham formalism of the density functional theory~\cite{hohenberg1964inhomogeneous,kohn1965self} is the {\it de facto} standard for performing the
electronic structure calculation of the atoms, molecules, solids, and clusters. While the theory is exact, the accuracy of
DFT depends on the approximations of the exchange-correlation (XC) functionals having all the many-body effects.
The development of efficient yet accurate XC functionals is an emerging topic in DFT for the last couple of decades and continues to be the same. 
Encouraging approximations have been  proposed in different levels, such as local density approximation
(LDA)~\cite{perdew1981self}, generalized gradient approximation (GGA)~\cite{perdew1996generalized,armiento2005functional,wu2006more,zhao2008construction,
PhysRevLett.100.136406,constantin2010communication,PhysRevB.79.075126,fabiano2014global,constantin2011correlation,doi:10.1021/ct200510s,constantin2016simple,
constantin2016semiclassical,cancio2018fitting,chiodo2012nonuniform}, meta-GGA~\cite{becke1989becke,voorhish1998novel,zhao2006new,perdew1999accurate,
tao2003climbing,perdew2009workhorse,constantin2012semiclassical,constantin2013metagga,sun2015strongly,ruzsinszky2012metagga,tao2016accurate,wang2017revised,
mezei2018simple,della2016kinetic,furness2019enhancing,sun2012effect,sun2013semilocal,aschebrock2019ultranonlocality,jana2019improving,constantin2016semilocal,patra2019relevance}, 
(screened-)hybrid density functional~\cite{heyd2003hybrid,paier2006screened,jana2020screened,staroverov2003comparative}, and
double-hybrid density functionals~\cite{goerigk2011efficient}, to improve the electronic structure
calculations of solids and quantum chemical systems. However, within the different rungs of the approximations, the semilocal form of the XC
approximations (LDA, GGA, and meta-GGA), proposed by satisfying exact constraints, are most common for the solid-state physics and quantum chemistry.
In the semilocal levels, the meta-GGA functionals are the advanced and
(more) accurate one for the solid-state and chemical calculations~\cite{hao2013performane,mo2017performance,haas2009calculation,
tran2016rungs,mo2017assessment,sun2011self,csonka2009assessment,jana2018assessment,jana2018assessing,patra2017properties,sun2013density,
patra2020electronic,patra2019efficient} which is written as, 
\begin{equation}
 E_{xc}^{meta-GGA}=\int~\rho({\bf{r}})\epsilon_{x}^{LDA}F_{xc}^{meta-GGA}(\rho,\nabla\rho,\nabla^2\rho,\tau)~,
\end{equation}
where, $\epsilon_{x}^{LDA}$ is the LDA exchange functional, $\rho,\nabla\rho,\nabla^2\rho,\tau$ are the density, gradient of 
density, Laplacian of density, and KS kinetic energy density, respectively. Due to the dependence of the KS kinetic energy and other 
built-in ingredients, the meta-GGA functionals recognize the single, overlap, and slowly varying density region in a much better
way than GGA does ~\cite{sun2013density}. However, in DFT it is a common
practice to construct more accurate semilocal XC by getting rid of the deficiencies of the proposed XC functionals.

Recent advances in the development of the semilocal functionals demonstrate that more accurate density functionals can be
proposed by satisfying as much as quantum mechanical constraints. One of the most important development came through the strongly
constrained and appropriately normed (SCAN)~\cite{sun2015strongly} meta-GGA functional, which is quite an accurate functionals for the diverse
nature of the solid-state and molecular properties~\cite{sun2016accurate}. Several modifications of the SCAN functional are also
proposed. Besides the SCAN functional, the recently proposed Tao-Mo (TM)~\cite{tao2016accurate} functional is also showing promising
performance for finite and extended systems~\cite{mo2017assessment,mo2017performance,jana2018assessment,jana2018assessing}. Besides, 
the built-in exchange hole of the TM functional extended to construct the range-separated hybrid density functionals for 
finite and extended systems~\cite{patra2018long,jana2018metagga,jana2019screened,jana2018many}. 
Also, a revision of the TM functionals (revTM) is proposed very recently~\cite{jana2019improving}. Importantly, the TM based functionals correctly satisfy two important paradigms, 
one- or two-electron limit, important for quantum chemistry and the slowly varying fourth-order density gradient approximation, relevant to condensed matter physics. 
However, both the TM and revTM functionals suffer from the order-of-limit problem anomaly, which is an important limitation and degrades 
its performance for the transition pressure solids~\cite{furness2020examining}. The order-of-limit problem of the meta-GGA functionals come 
from the iso-orbital indicator $z=\tau^{W}/\tau=1/(1+(3/5)(\alpha/p))$, where $\alpha=\frac{\tau-\tau^W}{\tau^{unif}}$ is another iso-orbital indicator, 
also known as Pauli kinetic energy 
density. Here, $p(=s^2=\frac{|\nabla\rho|^2}{((4k_F\rho)^2)})$, $\tau$, $\tau^W$, and $\tau^{unif}$ are the square of the reduced density gradient $s$, 
KS, von Weizsacker, and uniform kinetic energy density, respectively. In the limit of vanishing $\alpha$ and $p$ limit the order-of-limit
problem occurs as~\cite{perdew2004metagga,ruzsinszky2012metagga},
\begin{equation}
 \lim_{p\rightarrow 0}\big[\lim_{\alpha\rightarrow 0}\frac{1}{1+\frac{3}{5}\frac{\alpha}{p}}\big]=1.
\end{equation}
while 
\begin{equation}
 \lim_{\alpha\rightarrow 0}\big [\lim_{p\rightarrow 0}\frac{1}{1+\frac{3}{5}\frac{\alpha}{p}}\big]=0~.
\end{equation}
In the main paper of the TM functional~\cite{tao2016accurate}, the problem has been overlooked with statement 
 ``this only happens near a nucleus''. However, it is shown recently~\cite{furness2019enhancing} 
that the existence of this problem deteriorates the functional performance for the transition pressure of solids. Although the order-of-limit 
problem does not seem to be an important restriction for non-covalent ($\alpha>>1$) and slowly varying density region ($\alpha\approx 1$), but 
it seems to be a significant limitation for the center of the single-bonded region formed in molecules 
or solids. Also, during the change of the phases of two solids, where the formation of bonding and energy differences are important. 
Hence, during the meta-GGA functional construction, the order-of-limit problem must be taken into account. 
Note that TPSS and its revised version (revTPSS) also suffer from the order-of-limit anomaly, which is resolved by regularizing the functionals known as the regTPSS functional~\cite{ruzsinszky2012metagga}. Recent meta-GGAs like meta-GGA made simple (MS1, MS2, 
and MVS) and SCAN functionals do not have any order-of-limit problem because those are constructed using meta-GGA ingredient $\alpha$.

To find a way to resolve the order-of-limit problem of the TM functional (which, we named as regularized Tao-Mo (regTM) functional), 
in this paper, we propose a slightly different form of the iso-orbital indicator $z$.
Our resolution is described by arranging the paper as follows: In the next section, we will briefly describe the TM functional 
and its order-of-limit problem. A possible way to resolve the order-of-limit problem of the TM functional is also discussed in 
that section. Following this, we demonstrate the functional performance by assessing it for general purpose solids, molecules, 
and transition pressure problems.

\section{Theory}
To start, we first consider the functional form of the Tao-Mo (TM) functional and the underlying problems associated
with it. The exchange enhancement factor of the TM functional is given by~\cite{tao2016accurate},       
\begin{equation}
 F_x^{TM}(p,z,\alpha)=wF_x^{DME}+(1-w)F_x^{sc}~,
 \label{tmF}
\end{equation}
where
\begin{equation}
F_x^{DME}(p,\alpha)=\frac{1}{f^2}+\frac{7R}{9f^4}
\end{equation}
with $R=1+595(2\lambda-1)^2\frac{p}{54}-[z_3-3(\lambda^2-\lambda+1/2)(z_3-1-z_2/9)]$ and $f=[1+10(70y/27)+\beta y^2]^1/10$. 
Here, $z_2=\tau^{W}/\tau^{unif}=5p/3$, $z_3=\tau/\tau^{unif}=z_2+\alpha$, $y=(2\lambda-1)^2p$, $\tau^{VW}=|\nabla\rho|^2/(8\rho)$, 
and $\tau^{unif}=\frac{3}{10}(3\pi^2)^{2/3}\rho^{5/3}$. The slowly varying correction part of the enhancement factor given 
as~\cite{tao2016accurate},
\begin{eqnarray}
      F_x^{sc}(p,\alpha)=\bigg\{1+10\bigg[(\frac{10}{81}+\frac{50}{729}p)p+\frac{146}{2025}q^2\\ \nonumber
 -\frac{73}{405}q\frac{3z}{5}(1-z)\bigg] \bigg\}^{1/10}~.
\end{eqnarray}
where, $z=\tau^{W}/\tau=5p/(5p+3\alpha)$ and $w=(z^2+3z^3)/(1+z^3)^2$.
The order-of-limit problem of the TM exchange enhancement factor arises from the order of two limiting conditions $p\rightarrow 0$,
and $\alpha\rightarrow 0$. The discontinuity in the enhancement factor observed with,
\begin{equation}
 \lim_{\alpha\rightarrow 0}\big [\lim_{p\rightarrow 0}[F_x^{TM}(p,\alpha)]\big]=1.01372,
\end{equation}
and 
\begin{equation}
 \lim_{p\rightarrow 0}\big[\lim_{\alpha\rightarrow 0}[F_x^{TM}(p,\alpha)]\big]=1.1132.
\end{equation}
However, it is observed that none of the $F_x^{DME}(p,\alpha)$ and $F_x^{sc}(p,\alpha)$ face any order-of-limit problem,
\begin{equation}
  \lim_{p\rightarrow 0}\big[\lim_{\alpha\rightarrow 0}[F_x^{DME}(p,\alpha)]\big]=
 \lim_{\alpha\rightarrow 0}\big [\lim_{p\rightarrow 0}[F_x^{DME}(p,\alpha)]\big]=1.1132   
\end{equation}
and 
\begin{equation}
  \lim_{p\rightarrow 0}\big[\lim_{\alpha\rightarrow 0}[F_x^{sc}(p,\alpha)]\big]=
 \lim_{\alpha\rightarrow 0}\big [\lim_{p\rightarrow 0}[F_x^{sc}(p,\alpha)]\big]=1.01372.  
\end{equation}
The iso-orbital indicator $z$ present in the weight factor of Eq. \ref{tmF} is the root cause of the 
order-of-limit problem.

In search for appropriate weight factor, we propose modified iso-orbital indicator ($z'$) as,
\begin{equation}
z' = \frac{1}{1+(\frac{3}{5})\Big[\frac{\alpha}{p+f(\alpha,p)}\Big]}~,    
\end{equation}
where, the function $f(\alpha,p)$ is considered as proposed in ref.~\cite{ruzsinszky2012metagga} as $f(\alpha,p)
=\frac{(1-\alpha)^3}{(1+(d\alpha)^2)^{3/2}}e^{-cp}$ with $d=1.475$ and $c=3.0$. This choice of the iso-orbital indicator lifts 
the order-of-limit problem as,
\begin{equation}
  \lim_{p\rightarrow 0}\big[\lim_{\alpha\rightarrow 0}[z']\big]=
 \lim_{\alpha\rightarrow 0}\big [\lim_{p\rightarrow 0}[z']\big]=1~.   
\end{equation}
Note that any small positive definite quantity or real number instead of $f(\alpha,p)$ can remove the order-of-limit problem of $z'$. However, 
the present choice of $z'$ keeps the exchange enhancement factor close to that of the TM functional (except $s,  \alpha\to 0$), which is important. 
In ref.~\cite{ruzsinszky2012metagga}, the functional $f(\alpha,p)$ is considered to interpolation between $\alpha = 0$ and ordinary $\alpha$ values. 
But in the present case, $f(\alpha,p)$ is added to the square of the reduced density gradient ($p$) such that the function $\frac{\alpha}{p}$ becomes a 
finite number even for $p=0$, and a finite $\alpha$. We keep the original values of $c$ and $d$ as mentioned in 
ref.~\cite{ruzsinszky2012metagga}, since the order-of-limit is important only for $s\to 0$ and $\alpha\to 0$. Whereas, for other values of 
$s$ and $\alpha$, the regTM exchange enhancement factor matches smoothly with that of TM to keep the accuracy of the original functional.

\begin{figure}
    \includegraphics[width=3.2in,height=2.2in]{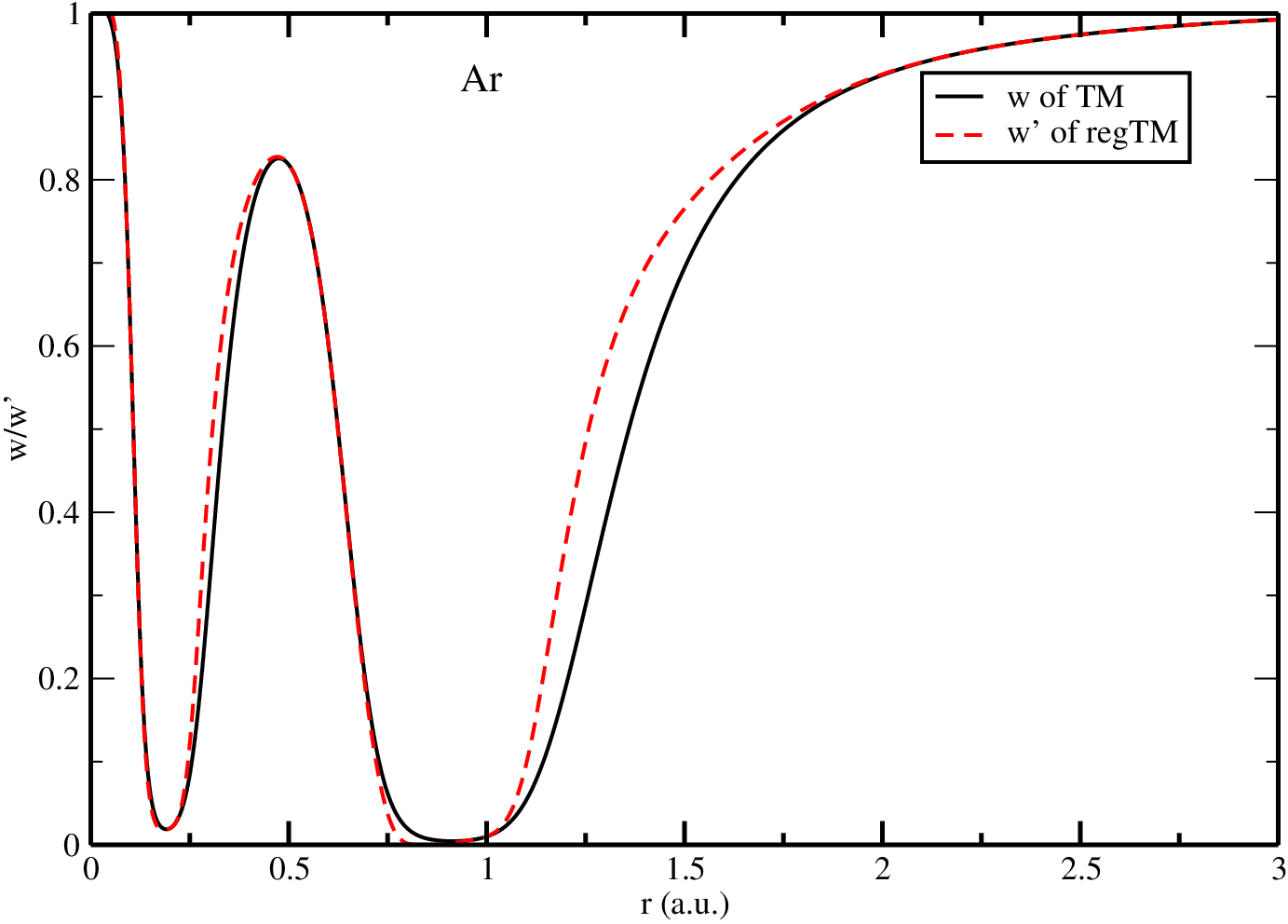}
    \caption{Shown is the $w$ of TM ($w'$ of regTM) for the Ar atom along the radial distance form the nucleus $r$.}
    \label{Ar-w}
\end{figure}

With the modified iso-orbital indicator the weight factor of the TM functional becomes,
\begin{equation}
w' = \frac{z'^2+3z'^3}{(1+z'^3)^2}~.    
\end{equation}

For one or two-electron singlet state, $\alpha=0$, that implies $z'=1$, $w'=1$, and $F_x^{regTM}=F_x^{DME}$. For
slowly varying density region, $\alpha\approx 1$, $f(\alpha,p)\approx 0$ and $w'$ is small. Hence, $F_x^{sc}$ is the 
dominating term that is necessary for solids. 
For non-covalent bonding, $\alpha>>1$, $f(\alpha,p)$ is small except for small
$s$. For example, $\alpha=10$, $f(10,p)$ is zero for $s>1.6$. The form of $f(\alpha,p)$ is such that $F_x^{regTM}$  matches
closely with that of the $F_x^{TM}$ for different $\alpha$ and $s$ values except for those two limiting values from which
order-of-limit occurs. Keeping the $F_x^{regTM}$ close to the $F_x^{TM}$ is important because $F_x^{TM}$ is a quite good
functional for weekly bonded systems and strongly bound solids, including the non-covalent interactions and layered
materials. Other forms of the $w'$ can be proposed based on the different iso-orbital indicators like $\alpha$ and $\beta$~\cite{furness2019enhancing}, 
but those may make the functional behavior quite different from $F_x^{TM}$, especially, for large $\alpha$, which is important for 
non-covalent systems~\cite{sun2013density,jana2020accurate}. 
For $\alpha=0$, $F_x^{TM}$ and $F_x^{regTM}$ reduce to $F_x^{DME}$ and for all values of $\alpha$. 
For comparison, in Fig~\ref{Ar-w}, we plot the $w$ and $w'$ for 
TM and regTM functional for Ar atom along the radial distance ($r$) from the nucleus. Both the $w$ and $w'$ match for the near 
nucleus and tail region ($\alpha\approx 1$). In the 
middle of the inter-shell region only slight difference is observed between these two curves indicating the consistency of the 
both $w$ and $w'$ by construction.

\begin{figure}
    \includegraphics[width=3.2in,height=2.2in]{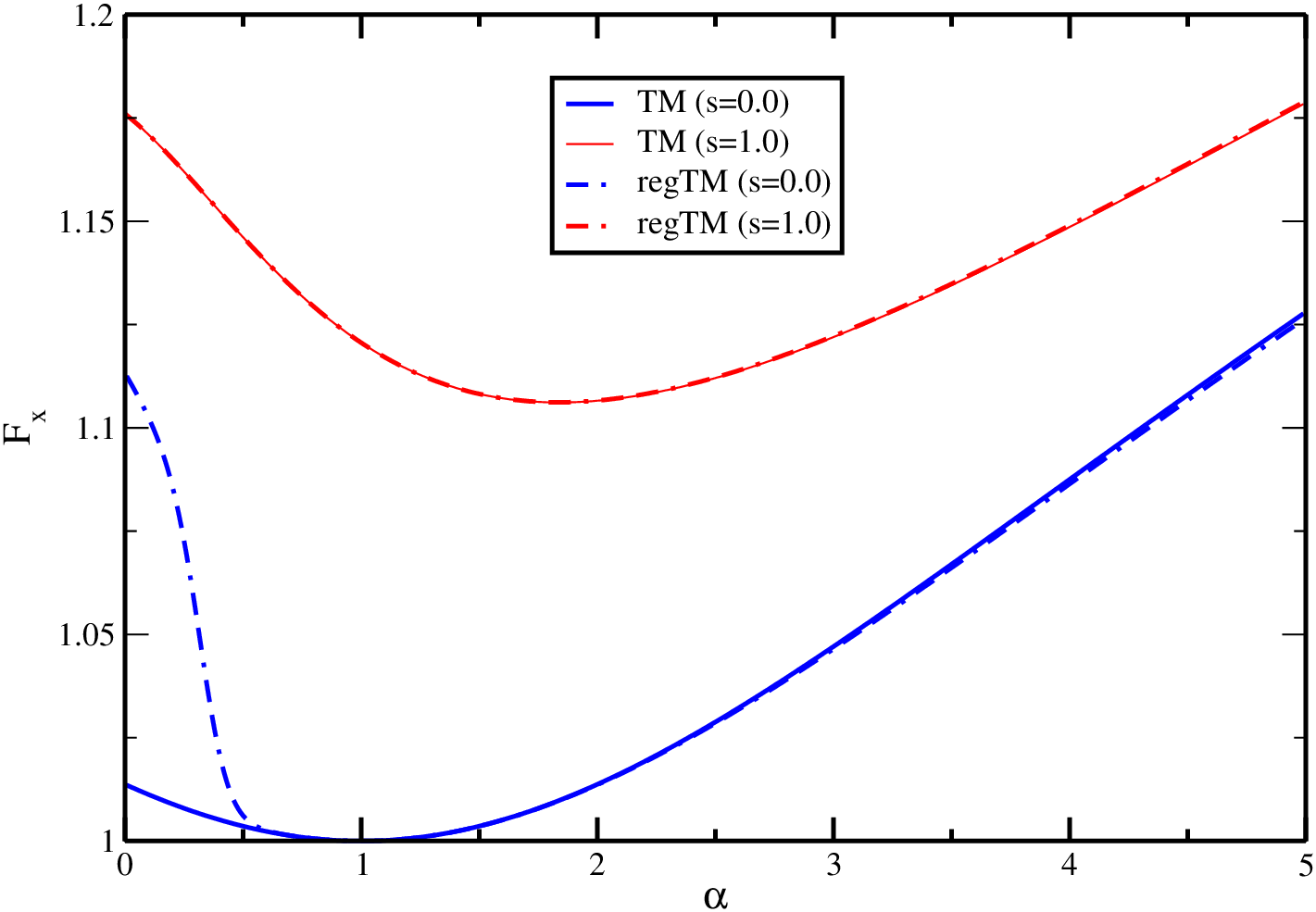}
    \caption{The difference in the enhancement factor of TM and regTM with the variation of $\alpha$ for particular 
    values of $s$ are shown. For $s=0$, the drastic difference at $\alpha =0$ is clearly from the correction of order-of-limit
    problem}
    \label{fx_alpha}
\end{figure}

Now, the use of $w'$ in Eq. \ref{tmF} lifts the order-of-limit problem and follows,  
\begin{equation}
 \lim_{p\rightarrow 0}\big[\lim_{\alpha\rightarrow 0}[F_x^{regTM}(p,\alpha)]\big]=
 \lim_{\alpha\rightarrow 0}\big [\lim_{p\rightarrow 0}[F_x^{regTM}(p,\alpha)]\big]=1.1132
\end{equation}
In Fig. \ref{fx_alpha},
we compare the variation in the enhancement factors of both TM and regTM exchange functionals with $\alpha$, and a
particular value of $s$. The order-of-limit correction is clear at $\alpha=0$ for $s=0$ curves. For other values of $\alpha$ and $s$,
the regTM exchange enhancement factor matches very closely with that of the TM functional. 

\begin{table}
\scriptsize
\begin{center}
\caption{\label{tab1} Tabulated are the mean absolute errors for molecular (main group thermochemistry, barrier heights and non-covalent interactions) and 
solid-state (equilibrium lattice constants, bulk moduli, and cohesive energies of a set of 29 bulk materials) tests. Best values within TM and regTM 
are marked with bold style. The zero-point an-harmonic 
expansion (ZPAE) corrected reference values for lattice constants, bulk moduli, and cohesive energies are taken from 
ref.~\cite{sun2011self,haas2009calculation,jana2019improving}. For LC20, BM20, and COH20 test sets are taken from refs.~\cite{sun2011self,ruzsinszky2012metagga}.
}
\begin{ruledtabular}
\begin{tabular}{llccccccccccccccccccccccc}
                  && TM &regTM   \\
\hline
\multicolumn{6}{c}{Molecular tests} \\[0.2 cm]
\multicolumn{6}{c}{Main group thermochemistry (kcal/mol)} \\
AE6&&4.5&{\bf 4.4}\\
G2/148&&6.5&{\bf 5.7}  \\
IP13 &&{\bf 3.17}&3.76  \\ 
EA13 &&3.79&{\bf 3.25}  \\   
PA8 &&{\bf 2.13}&5.13  \\
\multicolumn{6}{c}{Barrier heights (kcal/mol)} \\
BH6&&7.59&{\bf 5.47}\\
HTBH38&&7.25&{\bf 7.17}\\
NHTBH38&&8.86&{\bf 8.29}\\
\multicolumn{6}{c}{Non-covalent interactions (kcal/mol)} \\
HB6&&0.23&{\bf 0.10}  \\ 
DI6&&0.40&{\bf 0.30}\\
CT7&&2.87&{\bf 2.67}\\
PPS5&&0.74&{\bf 0.62}\\
WI7&&{\bf 0.04}&{\bf 0.04}\\
S22&&0.61&{\bf 0.55}\\
WATER27&&{\bf 1.44}& 1.53\\
\hline
TMAE&&3.34&{\bf 3.26}\\
\hline
\multicolumn{6}{c}{Solid-State tests} \\[0.2 cm]
\multicolumn{6}{c}{Lattice constants (\AA)} \\
simple metals&&0.051 &{\bf 0.044}      \\
transition metals&&{\bf 0.024}&0.026  \\
ionic solids&&0.039 &{\bf 0.037}        \\
semiconductors and insulators&&{\bf 0.015}&0.028        \\
total MAE&&{\bf 0.033}&0.034          \\
LC20&& {\bf 0.032}&0.033               \\
\multicolumn{6}{c}{bulk moduli (GPa)} \\
simple metals&&1.5  &{\bf 1.1}       \\
transition metals&&{\bf 9.9}&{\bf 9.9}   \\
ionic solids&&{\bf 4.2}&4.4         \\
semiconductors and insulators&&6.4&{\bf 6.1}         \\
total MAE &&5.5&{\bf 5.3}          \\
BM20&&4.2&{\bf 3.8}               \\
\multicolumn{6}{c}{cohesive energies (eV)} \\
simple metals&&0.274&{\bf 0.155}      \\
transition metals&&0.751&{\bf 0.570}  \\
ionic solids&&{\bf 0.069}&0.095         \\
semiconductors and insulators&&0.122&{\bf 0.078}     \\
total MAE&&0.304&{\bf 0.224}           \\
COH20&&0.282&{\bf 0.216}               \\
\end{tabular}
\end{ruledtabular}
\end{center}
\end{table}

Regarding correlation, we consider the regTPSS correlation energy functional~\cite{ruzsinszky2012metagga}. The regTPSS
correlation functional is also utilized with MS1, MS2, and MVS functionals, and it seems to be more suitable for
functionals proposed by removing the order of limit problem. The use of TM or TPSS correlation energy functional 
does not work well with the regTM functional as the obtained AE6 atomization energies are $\approx 8.0$ kcal/mol. The
only price the regTPSS correlation pays is that it is not one-electron self-correlation free, which is important for
molecules having many H atoms, such as water clusters. But as stated in ref.~\cite{ruzsinszky2012metagga}, molecular
reaction energies do not influence much by the atomic energy errors, and it depends on the error cancellation effects 
obtained from exchange and correlation. Also, the spin-independent of the XC functional for $0<\zeta<0.7$
~\cite{perdew2004metagga,sun2015strongly}, a constraint, important for improving the atomization energies, 
which TM XC functional respects well (see Fig. 2 of ref.~\cite{tao2016accurate}). 
    
\section{Results and Discussions}

\begin{figure}
    \includegraphics[width=3.2in,height=2.2in]{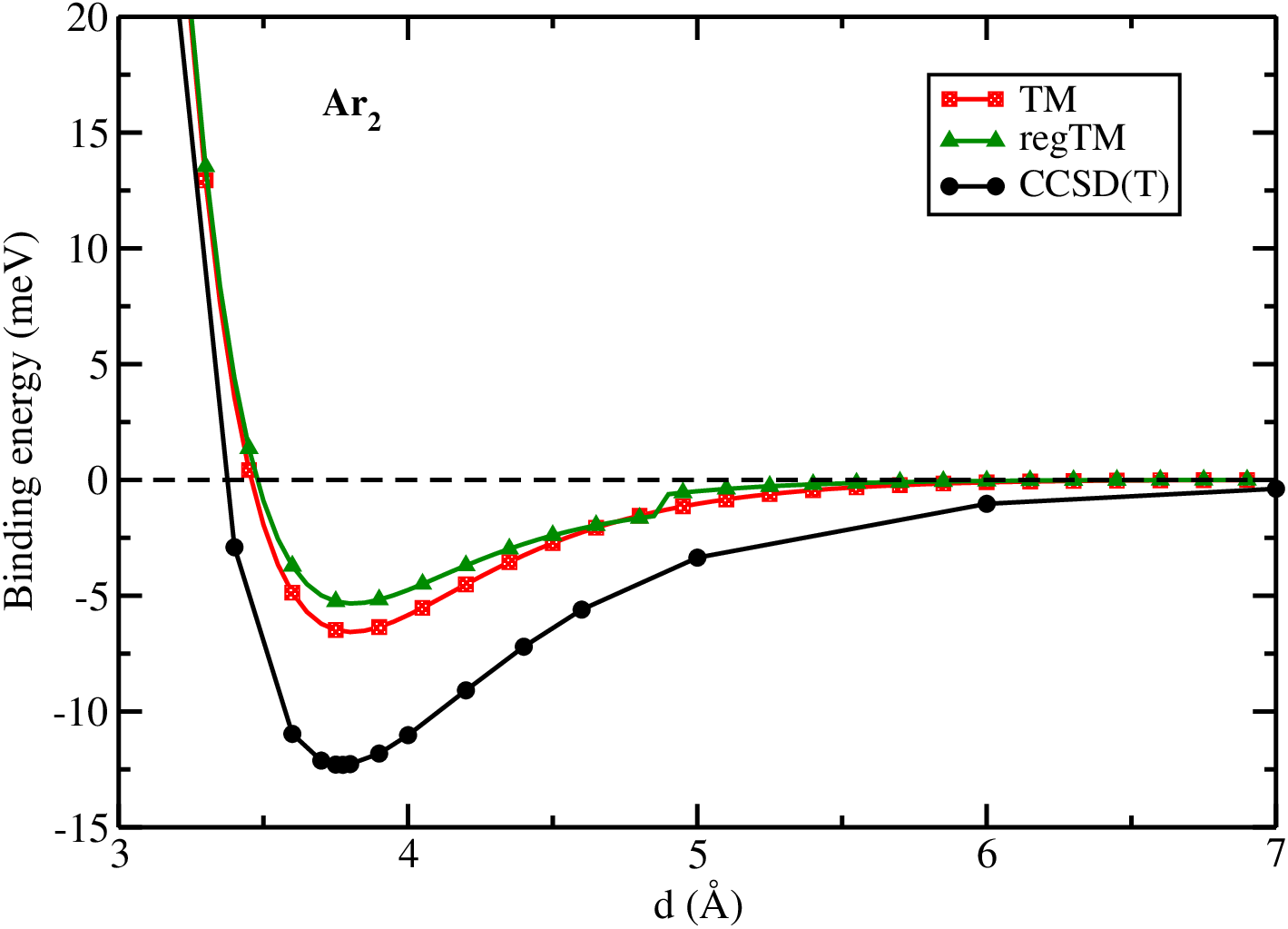}\\
    \vspace{2 cm}
    \includegraphics[width=3.2in,height=2.2in]{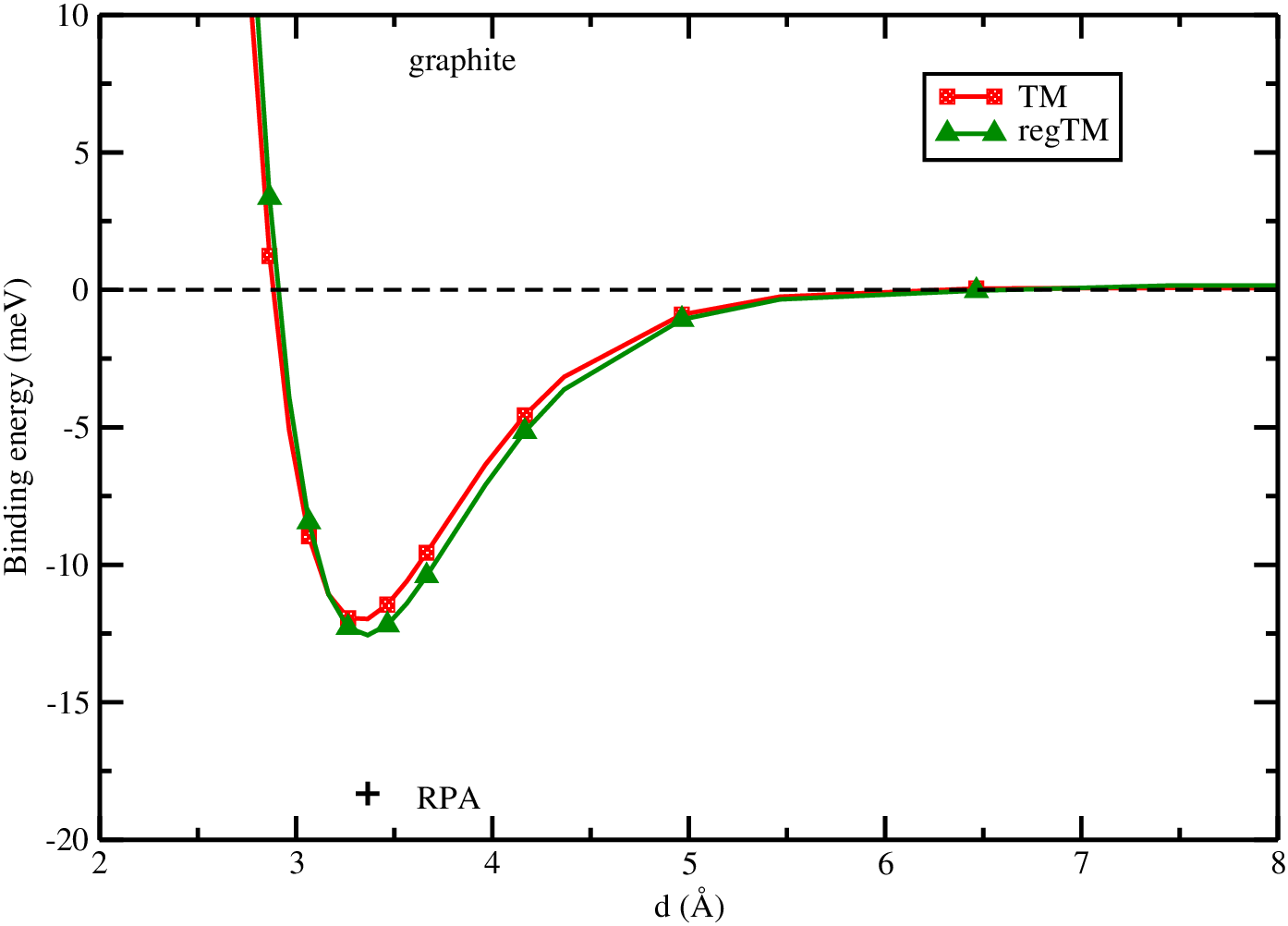}
    \caption{The binding energy curve of the Ar$_2$ dimer (upper panel) and graphite (lower panel) as obtained from TM and regTM functionals. The CCSD(T) 
    and RPA values are taken from ref.~\cite{konrad2005accurate,slavicek2003state} and ref.~\cite{bjorkman2012vanderwaals} respectively.}
    \label{binding-plot}
\end{figure}

\subsection{General assessment}

To assess the regTM functional performance along with TM functional, we consider general-purpose quantum
chemical and solid-state test sets. For quantum chemistry, the Minnesota 2.0~\cite{quest2014peverati} is considered except
G2/148 (atomization energies of $148$ molecules), S22 (22 non-covalent interactions), and WATER27 test sets~\cite{WATER27JCTC2009,WATER27JCTC2017}. The G2/148 test set is
considered from ref.~\cite{curtiss1997assessment}, whereas geometries of S22 and WATER27 are taken from GMTKN55~\cite{goerigk2017look}. Overall,
the test set is divided into the main group thermochemistry, barrier heights, and non-covalent interactions. The 
thermochemistry group consists of (1) AE6 - atomization energies of $6$ molecules~\cite{lynch2004small}, (2) G2/148 -
atomization energies of $148$ molecules~\cite{curtiss1997assessment}, (3) EA13 - 13 electron 
affinities~\cite{quest2014peverati}, (4) IP13 - 13 ionization potentials~\cite{quest2014peverati}, and (5) PA8 - 8 proton
affinities~\cite{quest2014peverati}. The barrier height test set consists with: (1) BH6 -  6 barrier
heights~\cite{lynch2004small}, (2) HTBH38 - 38 hydrogen barrier heights~\cite{zhao2005benchmark}, and  
(3) NHTBH38 - 38 non-hydrogen barrier heights~\cite{zhao2005benchmark}. For the non-covalent group we consider: (1) HB6 -
6 hydrogen bond test set~\cite{zhao2005design}, (2) DI6 - 6 dipole interactions~\cite{zhao2005design}, (3) CT7 - 7 charge
transfer molecules~\cite{zhao2005design}, (4) PPS5 - binding energies of five $\pi-\pi$ stacking complexes~\cite{zhao2005design}, (5) WI7 - 7 weekly interaction complexes~\cite{zhao2005design}, and (4) S22 - 22 non-covalent interaction molecules including H-bond, dispersion 
interactions, and mixed bonds~\cite{S22test2006,MarshallJCP2011}, and (5) WATER27 - 27 water cluster binding
energies~\cite{WATER27JCTC2009,WATER27JCTC2017}.

\begin{figure*}
    \includegraphics[width=3.0in,height=2.0in]{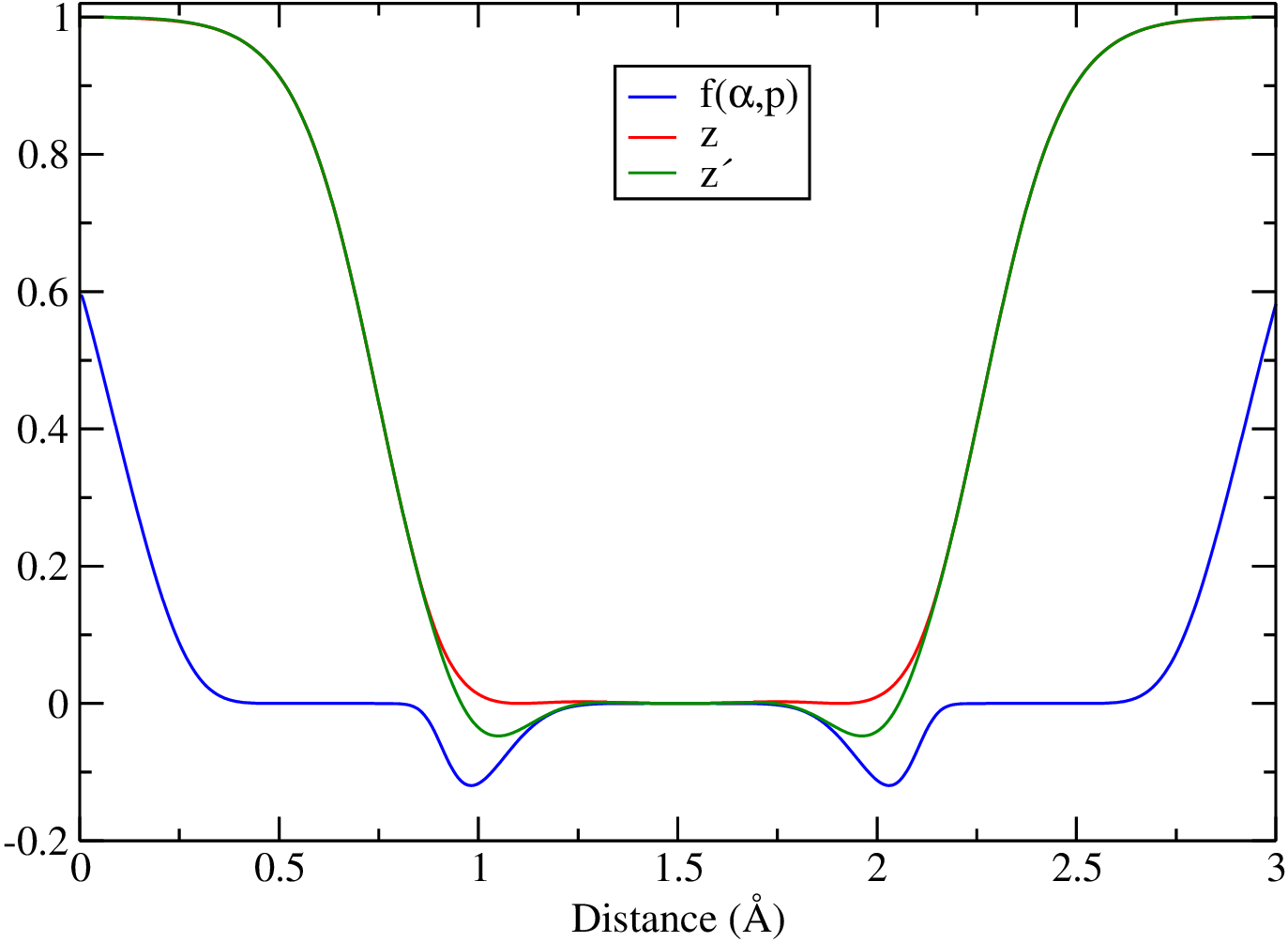}
    \hspace{1.0 cm}
    \includegraphics[width=3.0in,height=2.0in]{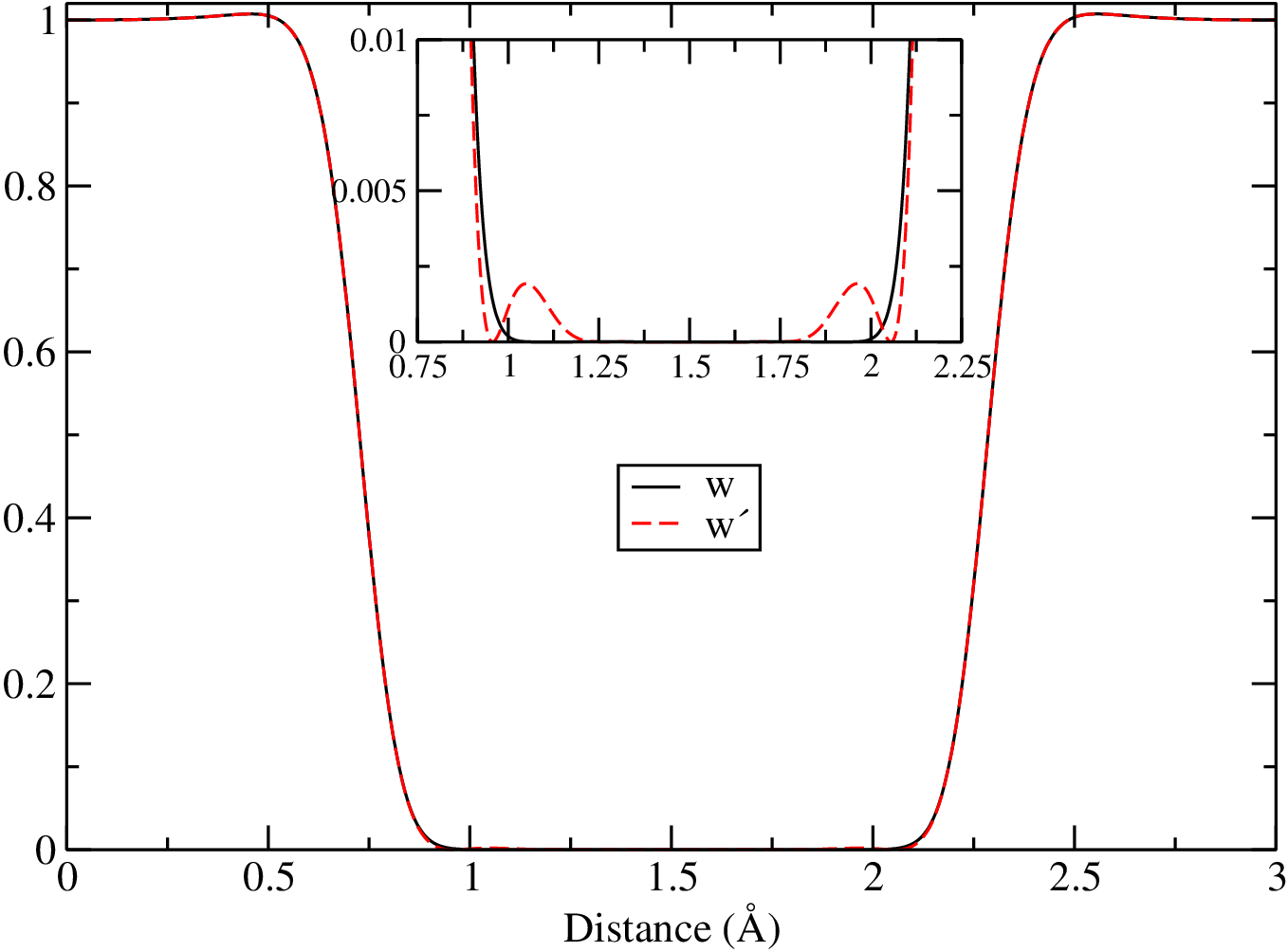}
    \caption{Shown are the different meta-GGA ingredients and interpolation functions for Li from the atom at (0, 0, 0) to (1/2, 1/2, 1/2). Left panel showing 
    $f(\alpha,p)$, $z$, and $z'$. Right panel showing $w$ and $w'$.}
    \label{fig4}
\end{figure*}

The performance of the TM and regTM functionals for the molecular test cases are summarized in Table~\ref{tab1}. Regarding the performance,  both functionals behave equivalently 
for main group thermochemistry except for PA8, for which the regTM worse than TM. For barrier height, regTM performs slightly better than TM both for the hydrogen and non-hydrogen 
transfer barrier heights. Importantly, the resolution of the order-of-limit problem seems important for the barrier height~\cite{ruzsinszky2012metagga}. Interesting results 
are also observed for H-bond and non-covalent interactions. Regarding the performance, regTM produces a slightly better result than TM for H-bonded, dipole interactions, charge 
transfer, and $\pi-\pi$ stacking complexes. The improvement for H-bonds for regTM functional is also reflected in the performance of the S22 test set, where regTM is overall better 
than TM functional for H-bonded and mixed complexes. We observe that for non-covalent interactions, especially for H-bonded systems, regTM binding energies are slightly lower 
than the TM functional, making binding energies closer to the experimental values. However, this moderate underestimation causes slightly worse performance of regTM for 
27 water clusters binding energies than TM functional. This behavior of the regTM functional for non-covalent interaction may happen because $F_x^{regTM}$ slightly enhanced 
than $F_x^{TM}$ (coming from $f(\alpha,p)$ function) for $\alpha>>1$, important for the non-covalent interaction. Also, one can not rule out the lack of one-electron self-interaction 
free correlation for regTM, which may also be responsible for the error cancellation between exchange and correlation.

\begin{figure*}
    \includegraphics[width=3.0in,height=2.0in]{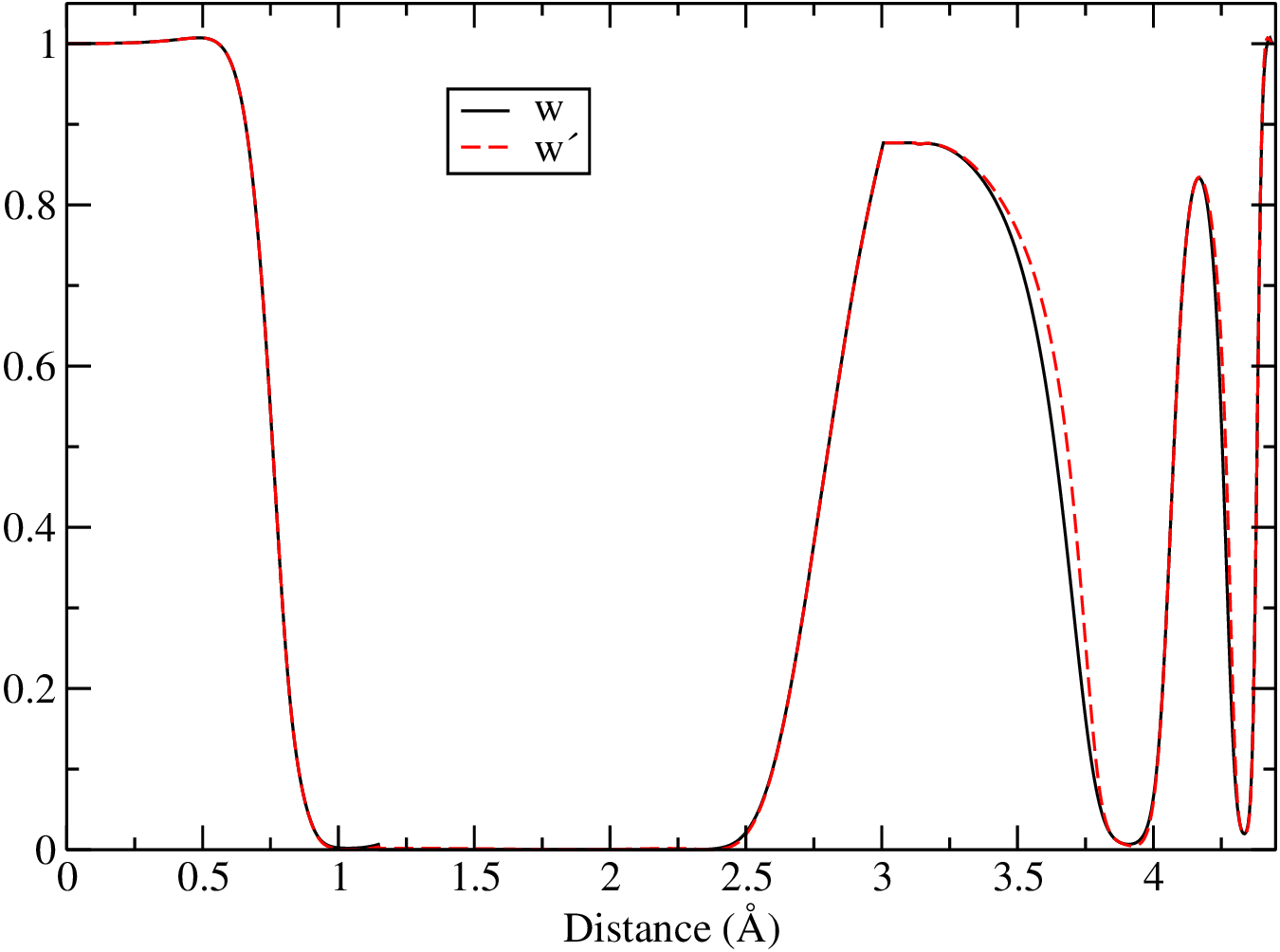}    
    \hspace{1.0 cm}
    \includegraphics[width=3.0in,height=2.0in]{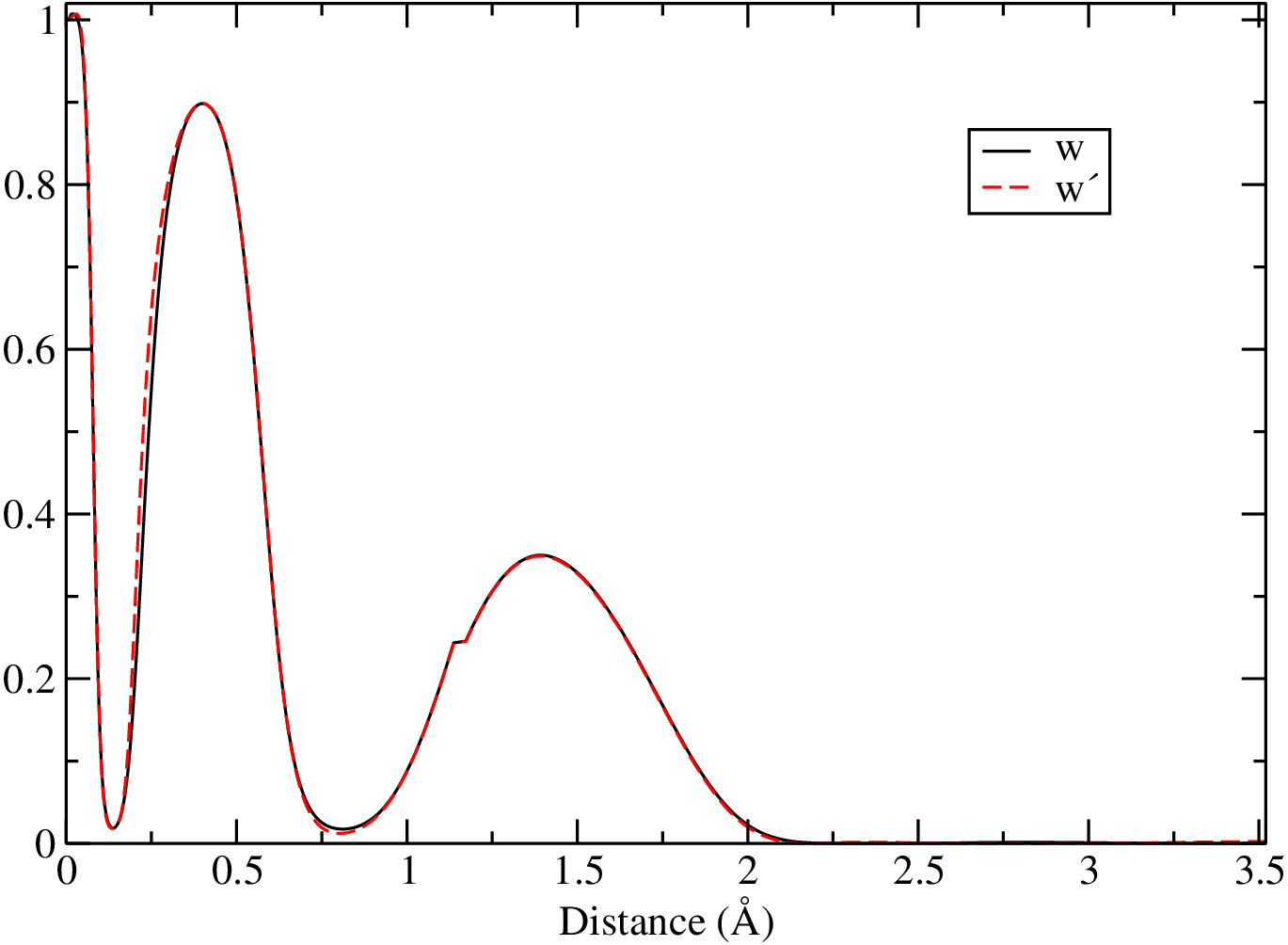}    
    \caption{Shown is the interpolation function $w$ and $w'$ for LiCl from the Li atom at (0, 0, 0) to the Cl atom at (1/2, 1/2, 1/2) (left panel) and 
    Si from the atom at (1/8, 1/8, 1/8) to (1/2, 1/2, 1/2) (right panel).}
    \label{fig5}
\end{figure*}

\begingroup
\begin{table*}
\scriptsize
\caption{Tabulated are the phase transition pressure ($P_t$) (in GPa), and energy difference ($\Delta E_e$) (in eV/functional) 
of  highly symmetric phases using TM and regTM functionals. For reference comparison, the ($\Delta E_e$) is compared with  SCAN/ ACSOSEX from ref.~\cite{sengupta2018from}.
All values are without temperature corrections.}
\begin{ruledtabular}
\begin{tabular}{lcccccccccccccccccccccccccccccccccc}

	        &\multicolumn{3}{c}{$P_t$}    &\multicolumn{3}{c}{$\Delta E_e$} 		\\ 	
Solids          & Expt.$^a$     &TM     &regTM       &SCAN/ACSOSEX$^a$  &TM     &regTM     \\ \hline
Si&12.0   &3.9    &14.5       &0.417/0.328    &0.246 &0.415     \\[0.05 cm]
Ge&10.6   &6.7    &7.5        &0.265/0.280    &0.365  &0.394      \\[0.05 cm]
SiC &100.0  &52.5   &66.9       &1.631/1.599	&1.227  &1.548      \\[0.05 cm]
GaAs& 15.0   & 8.2   &14.9       &0.825/0.978    &0.659  &0.728      \\[0.05 cm]
Pb    &14.0   & 10.0  &15.08      &0.015/0.027    &0.007  &0.024        \\[0.05 cm]
C     &3.7    &2.17   &4.81       &0.088/0.130    &0.040  &0.086       \\[0.05 cm]
BN  &5.0    &-1.2   &1.11       &0.105/0.048    &0.044  &0.042       
\label{tabpt}
\end{tabular}
\begin{tablenotes}
            \item[]a) See ref.~\cite{sengupta2018from} and all references therein.
            \end{tablenotes}
\end{ruledtabular}
\end{table*}
\endgroup


Next, we turn into the solid-state performance of the functionals. For solids the recovery of the slowly varying fourth-order gradient approximation of exchange is important. Both the 
TM and regTM functional recover the slowly varying fourth-order gradient approximation correctly. In Table~\ref{tab1}, we have shown the MAE in lattice constants, bulk moduli, and 
cohesive energies of $29$ bulk solids (compiled in ref.~\cite{constantin2012semiclassical}) as obtained from TM and regTM functionals. For lattice constants the zero-point an-harmonic 
expansion (ZPAE) corrected reference values are taken from ref.~\cite{sun2011self,haas2009calculation,jana2019improving}. It is shown, for accurate functional assessment, one should consider the ZPAE 
corrected lattice constant values~\cite{furness2020examining}. Regarding the lattice constants performance, TM and regTM perform almost similarly. Both functionals underestimate the 
lattice constants for metals and ionic solids. Note that solids like Li, Na, and K are also known as ``soft matter", for which the lattice constants is influenced by the short-range 
part of the vdW interaction~\cite{tao2010long}. Although, semilocal density functionals like MS1, MS2, MVS, and SCAN functionals  
include some amount of short-range part of the vdW interaction~\cite{sun2013density,sun2015strongly}, none of the TM and regTM 
functionals incorporate the short-range part of the vdW interaction and their good results solely depends on the error cancellation between 
exchange and correlation. Note that the correlation part of the TM functional proposed from modified TPSS correlation, which incorporate (i) one-electron self-interaction free correlation, (ii) correct slowly varying density 
limit of the correlation and local density linear response of the uniform electron gas limit, and (iii) spin independent of the XC functional for $0<\zeta<1$~\cite{tao2016accurate}.  
The constraint (ii) is only important for the solids, where the error cancellation between XC and restoration of good property of LSDA is important. In this respect the regTPSS correlation 
functional, used also in regTM functional,  keeps the correct formal properties of the solids i.e., correct slowly varying density limit of the correlation.
However, only differences in the TM and regTM construction from the lattice constants performance come from error cancellation between exchange and 
correlation which seems to be a bit better for TM functional, especially, for semiconductor and a few metals. Nevertheless, the performance of the regTM 
functional can be improved by incorporating a suitable meta-GGA ingredients dependent correlation energy functional compatible with the exchange 
or by including more exact constraints in the functional form. 
Overall, for the LC29 test set, we observe MAE of the 0.036 \AA~ from regTM functional compared to the 0.033 \AA~as obtained from TM functional.

To remark on the functional performance for bulk solids, in Fig.~\ref{fig4} we have shown the variation of $f(\alpha,p)$, $z$, $z'$, $w$ and $w'$ for the bulk solid Li 
(simple metal). The right panel of Fig.~\ref{fig4} shows that inside the bulk solid both the $w$ and $w'$ becomes one, while outside bulk both decay to zero, indicating 
the  likeness of the both interpolation parameters. However, a closer look (shown in left panel of Fig.~\ref{fig4}) suggests that for the overlap of 
the closed cell towards the valance region, $f(\alpha,p)$ and $z'$ slightly oscillatory (also $w'$), which also slightly shrink the lattice constant of Li for regTM. This slightly different behavior
of $w'$ purely coming form $f(\alpha,p)$ which depends on $\alpha$. We also observe similar behavior for LiCl (ionic solid) and Si (semiconductor) 
(shown in Fig.~\ref{fig5}). However, the similar behavior of $w$ and $w'$ indicate the judicious choice of $w'$. Overall, it correctly recovers $w$ 
throughout the range of the slowly varying bulk solids and rapidly or moderately varying atomic region.

Next, we focus on the bulk moduli of the solids. The bulk moduli are obtained from the equation of state fitting of the energy versus volume curve of the unit cell with the 
third-order Birch-Murnaghan isothermal equation of state. The volume of the unit cell is varied in the range $V_0\pm 5\%$, where $V_0$ is the equilibrium volume. This test depends on 
the accuracy of the geometries as predicted by the semilocal functional. As both the TM and regTM are quite good in predicting the geometries, the overall 
mean absolute errors obtain from both the functionals are about $\leq 5.5$ GPa. For comparison, we also show the BM20 results which can be directly 
compared with the results of other functionals presented in refs.~\cite{ruzsinszky2012metagga,sun2011self}.

The cohesive energy is also an important property for solids and it is related to the thermodynamics of solids. 
However, good accuracy for the lattice constants does not guarantee simultaneous good accuracy for cohesive energies. 
For example in the semilocal level, PBEsol does 
not perform as well as its lattice constants prediction for the cohesive energies. In most cases, meta-GGA functionals performance is better than 
GGA for cohesive energies 
because of their dependence on the additional ingredient i.e., KS kinetic energy density. For the accurate cohesive energies, simultaneous good performance for valence densities 
(which is moderately and rapidly varying) of atoms together with the bulk densities (which is slowly varying) is required. The cohesive energy per/atom is calculated as,
\begin{equation}
 E_{coh}=E_{atom}-\frac{E_{bulk}}{N}~,
\end{equation}
where $E_{atom}$ is the atomic energy and $E_{bulk}$ is the bulk energy of the unit cell having $N$ atoms.

Our results show that cohesive energies of alkali metals and transition metals are the most challenging for meta-GGA, where GGA PBE generally 
performs better~\cite{schimka2013lattice,jana2018assessing,peter2019comparative}. For transition metal, the total charge density becomes the sum of the several 
one-electron orbitals because of the largely filled $d$ shell. Therefore, the meta-GGA ingredients play a significant role in describing the 
energetic of the alkali metals~\cite{schimka2013lattice}. From Table~\ref{tab1}, we observe that, for alkali metals and transition metals,  
regTM shows improvement over the TM. Improvement of regTM over TM is also observed for semiconductors. While for ionic solids, TM performs a bit better.
The improved performance of regTM probably due to the slightly improved atomization energies and oscillatory nature of the $w'$ 
(and/or exchange enhancement factor) in the valance band of solids and core of the isolated atoms. Overall, regTM describes slightly better the 
slowly varying density and rapidly or moderately varying atomic region. Note that the SCAN functional also shows similar 
accuracy~\cite{jana2019improving} (or more accurate) for the $29$ test set presented in Table~\ref{tab1}. Nevertheless, one can not judge 
the improved performance of regTM from the modification in exchange only, the semilocal correction energy and long-range vdW interactions 
also play an important roles~\cite{tao2017screened,schimka2013lattice}

Lastly, in this section, we also focus on the TM and regTM performance of the binding energies of Ar$_2$ dimer and
graphite bi-layer. The Ar$_2$ dimer represents the non-covalent interaction and graphite bi-layer represents the layered materials. 
These two systems are often important to assess the quality of a functional for non-covalent interactions in molecular and solids level. The regTM binding energy 
curve a little bit up-shifted than TM functional for both the Ar$_2$ dimer and graphite bi-layer. This is probably due to the lack of the 
one-electron self-interaction free correlation. Nevertheless, regTM performs very closely to TM functional for these systems indicating 
the reliability ans closeness of $w$ and $w'$.    

\subsection{Structural phase transition}


Accurate prediction of the pressure-induced structural phase transition of solids form low-pressure phases (LP) to high-pressure phases (HP) 
having practical implication~\cite{vlasko2000direct,kang2014plasmonics,kato2003optical} and density functional semilocal and higher-order accurate 
wave functional methods are very successful in predicting the phase transition pressure~\cite{moll1995application,xiao2012structural,sengupta2018from,
kim2017origins,mellouhi2013structural,shahi2018accurate}. The accurate phase transition pressure depends on the accurate prediction of the both the 
equilibrium geometries and energy differences of the LP and HP phases. Regarding the various semilocal approaches, while LDA is good for structural 
properties, it tends to underestimate the energy difference of two phases~\cite{moll1995application,sengupta2018from}. The PBE GGA overestimates the 
volume and performs better than LDA for energy differences and phase transition pressure~\cite{moll1995application,sengupta2018from}. Recent advances 
in the development of the semilocal functional shows that the meta-GGA functionals like MS1, MS2, MVS and SCAN functionals performs better than PBE in 
predicting the phases transition pressure~\cite{xiao2012structural,sengupta2018from,shahi2018accurate}.

For meta-GGA functionals, the improved phase transition pressure is related to the artifact of the order-of-limit problem~\cite{ruzsinszky2012metagga,furness2020examining} 
which is related to the wrong energy difference between LP and HP phases. The meta-GGA having order-of-limit problem mostly underestimates (also overestimates in a few cases) 
the energy difference and hence predict the phase transition pressure wrongly, as it is shown for revTPSS functinal~\cite{ruzsinszky2012metagga}. The TM functional which 
also possesses the order-of-limit problem underestimates the phase transition pressure~\cite{furness2020examining}. As state in ref.~\cite{ruzsinszky2012metagga}, the order-of-limit 
problem is more severe for covalent bonded solids, where the $\alpha\approx 0$, and $s\approx 0$ often encounter around the critical bond point and shows the wrong energy difference 
between two phases of solids.

To illustrate the improvement of the phase transition pressure from regTM functional, in Table~\ref{tabpt}, the structural phase transition parameters as obtained from the TM and 
regTM functional are compared. We observe that the regTM yields phase transition pressure and energy differences close to the experimental one than the TM functional. 
For Si phase transition, regTM functional predicts improves considerably for energy difference and energy difference of two phases. Comparing the SCAN and ACSOSEX values 
from ref.~\cite{sengupta2018from}, the regTM is considerably close to that SCAN, indicating its improvement over TM upon eliminating the order-of-limit problem. A very 
similar tendency is observed for other structural phase transitions, where regTM improves considerably over TM functional. The regTM phase transition pressure of Si, 
GaAs, and Pb become very close to the experiment, where TM underestimates considerably. For those solids, the energy difference between the two phases is also close to that of 
the SCAN/ASCOSEX values~\cite{shahi2018accurate}. For cubic to hexagonal phase transition of BN, the aggrement of phase transition pressure for TM functional is very poor. It 
shows negative phase transition pressure, where regTM improves considerably over TM. However, we do not include the temperature corrections in our calculations, which as par 
shows about to improve the phase transition pressure considerably for BN~\cite{sengupta2018from}.

It is noteworthy to mention that accurate prediction of the phase transition pressure depends both 
the differences, the energy ($\Delta E_e$) and volume ($\Delta V_0$) differences. While both the functinal perform almost similarly for geometries, 
regTM performs better for energy differences. Therefore, it is clearly indicating that the elimination of the order-of-limit problem is important for the improvement of 
the structural phase transition properties from the semiconductor to metallic phases.


\begingroup
\begin{table}
\caption{Selective semiconductor band gaps (in eV) as obtained from regTM and TM functionals.}
\begin{ruledtabular}
\begin{tabular}{ccccccccccccccccccccccccccccccccccc}
Solids      &Expt.      & TM    & regTM      \\
\hline
C&5.5&4.08&5.08\\
Si&1.17&0.56&1.26\\
Ge&0.74&0.32&0.40\\
SiC&2.42&1.29&1.75\\
BP&2.4&1.20&1.94\\
GaAs&1.52&0.91&1.01
\label{bgap}
\end{tabular}
\end{ruledtabular}
\end{table}
\endgroup


\subsection{Connection to the band gap}


On eliminating the order-of-limit problem, the regTM functional also improves the semiconductor band gaps.
In Table~\ref{bgap} we have shown the band gaps of few selective solids for which a clear improvement of regTM over TM is  
evident. For diamond C, the regTM band gap increases by almost $1$ eV. Similarly, for Si, Ge, SiC, BP, and GaAs, we observe the 
improvement in the band gap of solids. However, this improvement is related to the slope of the exchange enhancement factor which is 
discussed in ref.~\cite{aschebrock2019ultranonlocality,patra2020electronic}. For regTM the slope $\partial F_x/\partial\alpha$ is more negative, which 
includes more derivative discontinuity ($\Delta_{xc}$) in the generalized KS scheme~\cite{aschebrock2019ultranonlocality}. This also 
happens for functionals like MS1, MS2, MVS, SCAN, MGGAC, and TASK functionals behavior. However, for regTM this happens only 
for $\alpha\approx 0$ and $s \approx 0$ region, relevant for the covalent bonded systems. For other regions, regTM exchange enhancement 
factor matches closely to that of the TM functional.

Next, we consider the relation between improvement to the phase transition pressure and the band gap, which is discussed in 
refs.~\cite{hennig2010phase,xiao2013testing}. In ref.~\cite{hennig2010phase}, it is argued that the underestimation of the phase transition pressure for Si for LDA and GGA may be related to the
band gap of solids. While higher order methods like GW and screened hybrid functional HSE06 enhance the density of state (DOS) near Fermi level and
describes the covalent bonding more conveniently which may responsible for the improved phase transition pressure for those methods
~\cite{hennig2010phase}. Contrary to ref.~\cite{hennig2010phase}, in ref.~\cite{xiao2013testing}, it is stated that the impact of the 
fundamental band gap may not be so important for the improvement in the phase transition pressure. However, in this case, the band gap improvement for regTM 
for semiconductors, especially for Si, indicates that the regTM transition pressures may be related to the improvement to the band 
gap in Si as suggested in ref.~\cite{hennig2010phase}.

\section{Conclusions}

The order-of-limit problem is an important limitation of the TM functional to predict the phase transition pressure
as shown in ref.~\cite{furness2020examining}. In this paper, a modified interpolation function of the Tao-Mo (TM) functional is proposed
which resolves the order-of-limit of the TM functional. Using the modified interpolation function, the proposed
functional correctly retains the accuracy of the parent functional for one- or two-electron limit, slowly varying
density correction, and non-covalent interacting systems. This is important because along with the resolution of
the order-of-limit problem, we retain the main functional accuracy for thermochemistry and solid-state physics which are
simultaneously important. It is shown that the phase transition pressure of the proposed functional is improved
corresponding to the TM functional.  As claimed in ref.~\cite{furness2020examining}, redesign the interpolation function as a function of
$\alpha$ may also resolve the order-of-limit problem of the TM functional, but in that case, one has to ensure simultaneous good accuracy of the parent functional. In these prospects, the present modification of the TM
functional to make it free from the order-of-limit problem seems quite suitable.   

Lastly, we conclude that the present modification of the iso-orbital indicator $z$ is simple and quite useful as it can be used further to construct meta-GGA functionals development. 
Along this line of construction, a one-electron self-interaction free correlation with the regTM exchange functional may also enhance
the performance of it for several thermochemical and solid-state structural properties. However, this can be further revisited in
future publications.

{\textbf{Computational details:}}~The molecular
calculations of the functionals are performed using
the developed version of Q-CHEM code~\cite{qchem} with def2-QZVP basis
set (for IP13 and WATER27 the def2-QZVPD basis set is
used) with 99 points radial grid and 590 points
angular Lebedev grid. Note that like both the TM and
regTM functionals are not sensitive on the choice
of the grid. Using the 350 points radial grid and 590
points angular Lebedev grid also we do not see any
difference in the potential energy curves for
non-bonded interactions. Hence, the choice of the 
more dense grid is not required and the present choice
of the grid is quite adequate for the energy
convergence.

All solid-state calculations are performed in plane
wave suite code {\it {Vienna ab initio simulation
package}} (VASP)~\cite{vasp1,vasp2,vasp3,vasp4,vasp5,vasp6,vasp7}. The lattice constants are performed by relaxing the 
volume and internal co-ordinates using conjugate gradient algorithm. For bulk calculations we used
20$\times$20$\times$20 $\Gamma$ centered $\bf{k}$
points with 800 eV energy cutoff. The spin polarized
atomic calculation for the cohesive energies are
performed with the orthorhombic box of size
of 23$\times$24$\times$25 \AA$^3$. To calculate the
bulk moduli and phase transition pressure third order Birch-Murnaghan equation of state~\cite{birch1947finite} is used.

\section*{Acknowledgements}
  
S.J. is grateful to the NISER for partial financial support. Q-CHEM and VASP simulations has been performed on {\it KALINGA} and {\it NISERDFT} 
high performance computing facility at NISER, Bhubaneswar. P.S. thanks Q-CHEM Inc. and developers for providing the source code.

\section{Data availability}

The details of the pseudopotential used in this calculations along with all the results are supplied in the 
supporting information.

\twocolumngrid
\bibliography{reference}
\bibliographystyle{apsrev4-1}

\end{document}